\documentclass[letterpaper]{mc2021}
\usepackage{tabls}
\usepackage{cites}
\usepackage{epsf}
\usepackage{appendix}
\usepackage{ragged2e}
\usepackage{subfigure,amsmath}
\usepackage{tikz}
\usepackage{verbatim}
\usetikzlibrary{intersections} \usetikzlibrary{shapes.geometric, arrows, positioning}

\tikzset{
  box/.style  = {draw,rectangle, minimum width=5cm, minimum height=1.2cm, text centered, text width=5cm, font=\Large},
  myarrow/.style = {line width=2mm, draw=blue!30, -triangle 60, fill=blue!30,postaction={draw, line width=4mm, shorten >=6mm, -}}
}
\usepackage[top=1in, bottom=1in, left=1in, right=1in]{geometry}
\usepackage{enumitem}
\setlist[itemize]{leftmargin=*}

\title{Frequency-Dependent Material Motion Benchmarks for Radiative Transfer}
\author{%
  \textbf{Ryan G.\ McClarren$^1$, N.A.\ Gentile$^2$}\\
  $^1$Dept.\ Aerospace and Mechanical Engineering, University of Notre Dame\\
  Notre Dame, Indiana, USA
\\
  $^2$L-38 Lawrence Livermore National Laboratory\\
    Livermore, California, USA \\ 
  \url{rmcclarr@nd.edu}, \url{gentile1@llnl.gov}
}
\newcommand{\authorHead}{McClarren and Gentile}
\newcommand{\shortTitle}{Frequency-Dependent Material Motion Benchmarks for Radiative Transfer}

\newcommand{\IL}{I_{\mathrm{L}} }
\newcommand{\nuL}{\nu_{\mathrm{L}} }

\newcommand{\DL}{D_{\mathrm{L}} }
\begin{document}
\maketitle
\justify 

\begin{abstract}
We present a general solution for the radiation intensity  in front of a purely absorbing slab moving toward an observer at constant speed and with a constant temperature. The solution is obtained by integrating the lab-frame radiation transport equation through the slab to the observer. We present comparisons between our benchmark and results from the Kull simulation code for an aluminum slab moving toward the observer at 2\% the speed-of-light. We demonstrate that ignoring certain material motion correction terms in the transport equation can lead to 20-80\% errors with the error magnitude growing as the frequency resolution is improved. Our results also indicate that our benchmark can identify potential errors in the implementation of material motion corrections.
\end{abstract}

\section{Radiation Spectrum in Front of a Moving Slab}
We are interested in solving the problem of the radiation spectrum in front of a slab that is moving at a fast speed that is, nevertheless, slow relative to the speed of light.  In our notation, the subscript L denotes a quantity in the laboratory reference frame, and the subscript F is for the fluid, or co-moving frame.
\begin{figure}[ht]
\begin{center}
\begin{tikzpicture}[scale=1.1]
    \draw [name path = rect1] (0,0) -- (2, 0) -- (2,4) node (topright) {} -- (0,4) node (topleft) {} -- (0,0);
    \coordinate (z0) at (0,0);
    \draw [ultra thick, name path = axes1] (z0) node [above right] {$z=0$} -- (12,0) node (z1) {};
    \path (topleft) ++(0,-1.5) coordinate (d);
    \path (z1) ++(0,-0.5) coordinate (e);
    \draw [->, name path=ray1] (d) -- (e);
    
    \path [name intersections={of = ray1 and axes1}];
  \coordinate (Zpos)  at (intersection-1);
    \draw (Zpos) node (Z) [circle, fill, scale=0.5] {};
    \draw (Zpos) node (Zlab) [below] {$Z$};

    \path [name intersections={of = ray1 and rect1}];
  \coordinate (eb)  at (intersection-1);
    \path (d)++(0,-0.15) coordinate (dbelow);
    \path (eb)++(0,-0.15) coordinate (ebelow);
    \draw [<->, color=blue] (dbelow) -- (ebelow) node [color = black, midway, circle, fill=white, yshift=-0.25] {$s$};
    
    \path (z0) ++(0,1) coordinate (L0);
    \path [name intersections={of = axes1 and rect1}];
  \coordinate (L1p)  at (intersection-1);
    \path (L1p) ++(0,1) coordinate (L1);
    
  \path [name path = circZ] (Zpos) circle (2);
    \path [name intersections={of = ray1 and circZ}];
  \coordinate (ic1)  at (intersection-1);
    \path [name intersections={of = axes1 and circZ}];
  \coordinate (ic2)  at (intersection-1);
  \draw [color=white] (ic1) -- (ic2)  node [midway, color=black, anchor=east] {$\theta$};
  \begin{scope}
\clip (Zpos) circle (2);
\clip (Zpos) -- (z0) -- (d) -- (Zpos);
\draw (Zpos) circle (1.99);
\end{scope}
  
  \draw [myarrow] (2,3.25) node [anchor=west,font = \footnotesize \sffamily] {slab motion} -- (4,3.25) ;
    
    \draw [<->, color=blue] (L0) -- (L1) node [color = black, midway,circle, fill=white] {$L$};
    \node[anchor=north west,text width=3cm, font = \footnotesize \sffamily] (note1) at (0.1,3.8) {
   Distance $s$ is\\ measured\\ along ray};
   \draw [color=white] (0,4.25) -- (2,4.25) node[midway,color=black,font = \footnotesize \sffamily] (note1) {slab};
    \node[anchor=north west,text width=3cm, font = \footnotesize \sffamily] (note1) at (5.5,2.75) {
  vacuum};
  \draw (Zpos) -- (d) node [midway, above,font = \footnotesize \sffamily] {ray};
\end{tikzpicture}
\end{center}\vspace{-2cm}
\caption{ {\bf Schematic of the problem: a slab of length $L$ moves in the $z$ direction toward
    the observation point at $z = Z$.  The solution $I(Z,\mu,\nuL,t)$  with $\mu = \cos \theta$ is computed by integrating along the ray over the length $s$, given by Eq.~\eqref{eq:s}.}}
\label{IntersectionDiagramSlab}
\end{figure}
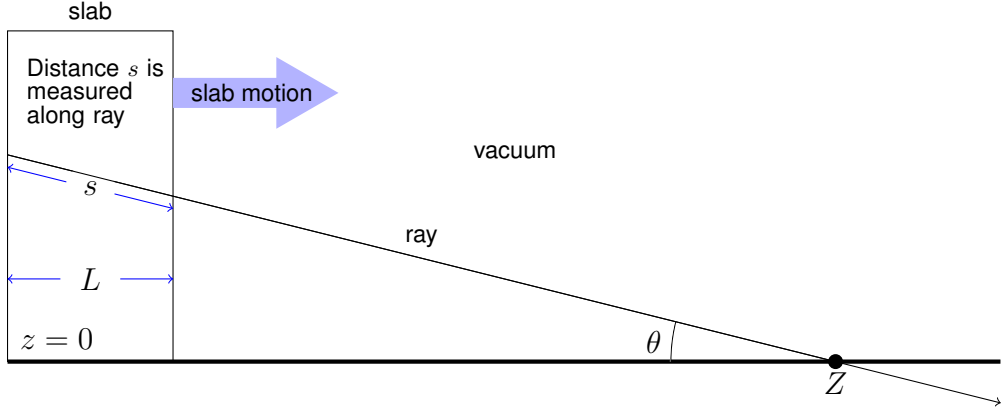

Considering the observation point $z=Z$, we  compute the solution $\IL(Z,\mu,\nuL,t)$. This problem is depicted in 
Fig.~\ref{IntersectionDiagramSlab}.  For a photon to reach position $Z$ at time $t_Z$ it will have its position $z(t)$ given by
\begin{equation}\label{eq:zpos}
z(t) = Z + \mu c (t-t_Z).
\end{equation}
Given that the back of the slab has position $z_b = v t$, a photon that leaves the back of the slab and reaches $Z$ at time $t_Z$ will be emitted at time $t_b$ that can be determined from the equation $z(t_b) = v t_b$ as
\begin{equation}\label{eq:tb}
t_\mathrm{b} = \left( \frac{\mu c t_Z - Z}{\mu c - v}\right)_+,
\end{equation}
where $(\cdot)_+$ gives the positive part of its argument (i.e., $(x)_+ = x$ if $x>0$, and $(x)_+ = 0$ otherwise). Similarly, a photon reaching $Z$ at time $t_Z$ emitted from the front of the slab at time $t_f$:
\begin{equation}\label{eq:tf}
t_\mathrm{f} = \left( \frac{L + \mu c t_Z - Z}{\mu c - v}\right)_+.
\end{equation}
Additionally, the length of the ray through the slab for a photon is then
\begin{equation}\label{eq:s}
s = c(t_f-t_b)
\end{equation}

Using these relations the solution $I(Z,\mu,\nuL,t_Z)$ 
is
\begin{equation}\label{eq:Izsol}
I(Z,\mu,\nuL,t_Z) = I_Z(\mu_L, {\nuL}) = \frac{B(\gamma {\DL} {\nuL}, T)}{(\gamma {\DL}) ^3}
\left[ 1 - \exp\left(-\gamma {\DL} \sigma_\mathrm{a,slab}(\gamma {\DL} {\nuL}) s \right) \right],
\end{equation}
which can be found by using an integrating factor.


\section{Analytic Results and Code Benchmarking}
\begin{figure}
\subfigure[$\kappa_g$]{\includegraphics[width=0.49\textwidth]{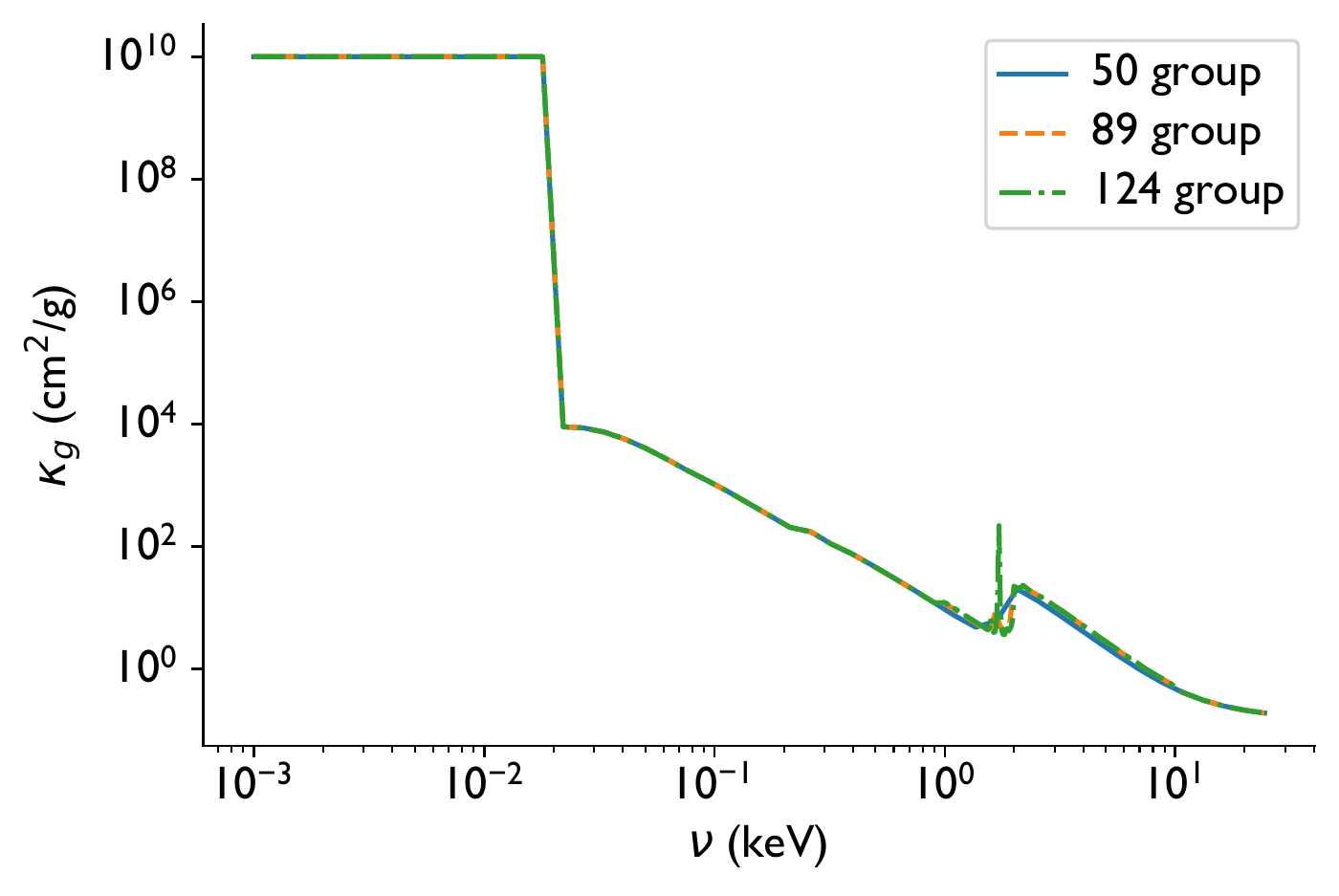}}
\subfigure[Detail near spectral lines]{\includegraphics[width=0.49\textwidth]{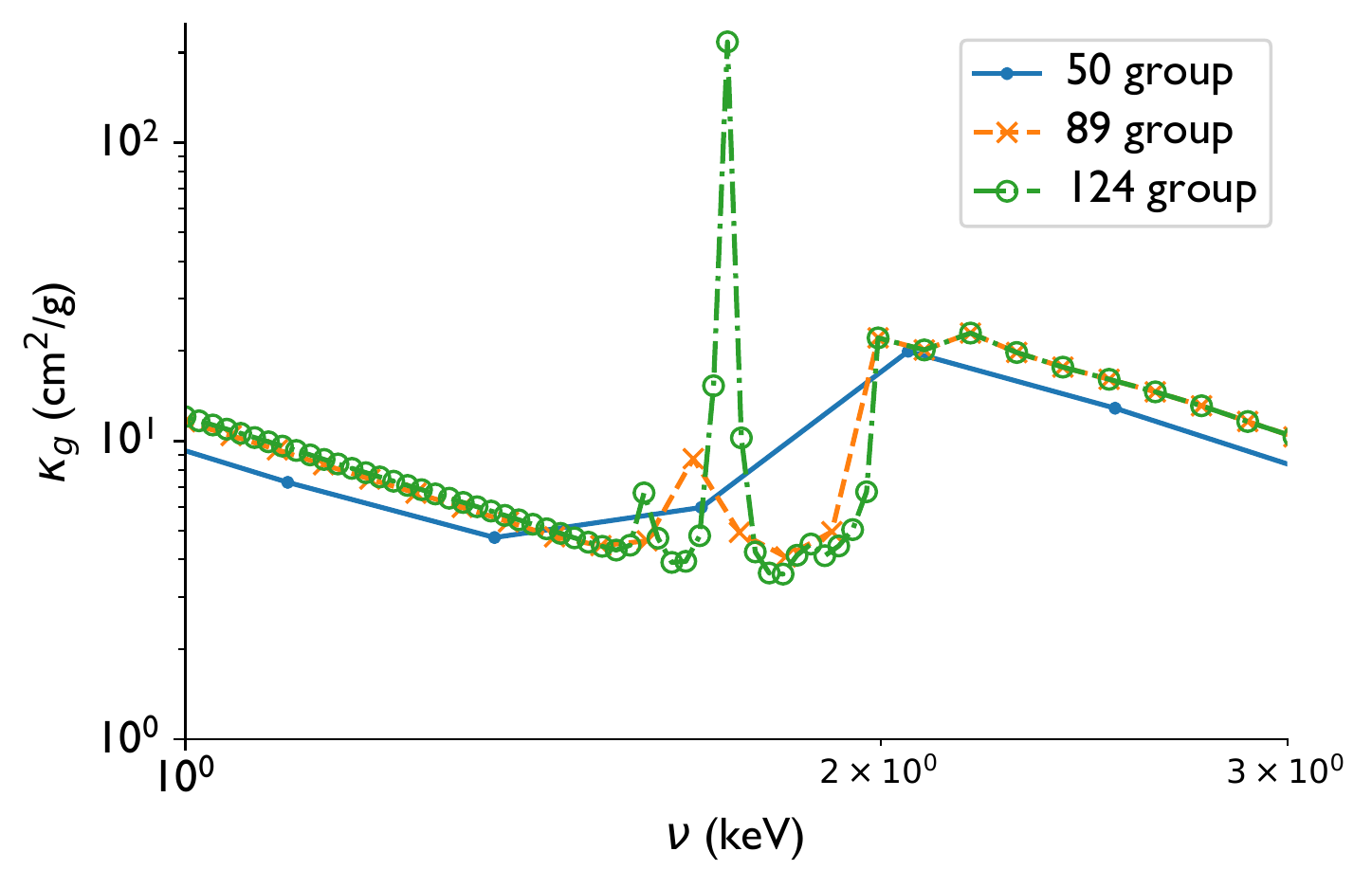}}
	\caption{ Group specific opacities, $\kappa_g$ in units of cm$^2$/g for the aluminum slab.}
	\label{fig:kappa}
\end{figure}

We consider a slab of aluminum with density of 0.1 g/cm$^{3}$, a thickness of $L=0.4$ cm., and a temperature of $T=1$ keV.  The slab is moving at a speed of 0.5994 cm/ns (about 2\% of the speed of light), a speed chosen based on a Mach 45 radiating shock benchmark \cite{lowrie2008radiative}. The observation point is $Z=12$ cm and time $t_Z = 10$ ns. We compute the multigroup radiation energy density based on three group structures:
\begin{itemize}
\item A coarse structure with 50 logarithmically spaced groups from 0.001 keV to 30 keV,
\item A medium structure of 89 total groups based on the coarse set where the extra groups are added between 1 and 10 keV,
\item A fine structure of 124 groups based on the medium set where the extra groups are added between 1 and 2 keV.
\end{itemize}
These group structures were chosen to successively capture the spectral lines in the 1 and 2 keV range. The opacities were provided by the TOPS opacity database \cite{magee1995atomic} and are shown in Figure \ref{fig:kappa}.

Our benchmark solution was computed by using Mathematica to compute the solutions.  We can also investigate the effects of material motion corrections (MMC) by computing the solution for a stationary slab (i.e., $v=0$), and when the frequencies in the opacity and blackbody emission terms are not Doppler shifted by $\gamma \DL$ (we call this the no $\nu$ Doppler solution). Even though in our problem $v/c \approx 0.02$, we will be able to show that ignoring MMC leads to errors on the order of 80\% for this problem.

We can also compare the benchmark to solutions computed by the implicit Monte Carlo method \cite{FC,smedley2015asymptotic} in Kull \cite{Kull}. We can verify that the MMC is correctly implemented in Kull by comparing its solution to the benchmark.  We can also compare Kull solutions with different MMC terms turned off, for example the Doppler shift of $\nu$ and show that the benchmark can elucidate that these terms are missing.

\begin{figure}
\subfigure[Benchmark and KULL Result]{\includegraphics[width=0.33\textwidth]{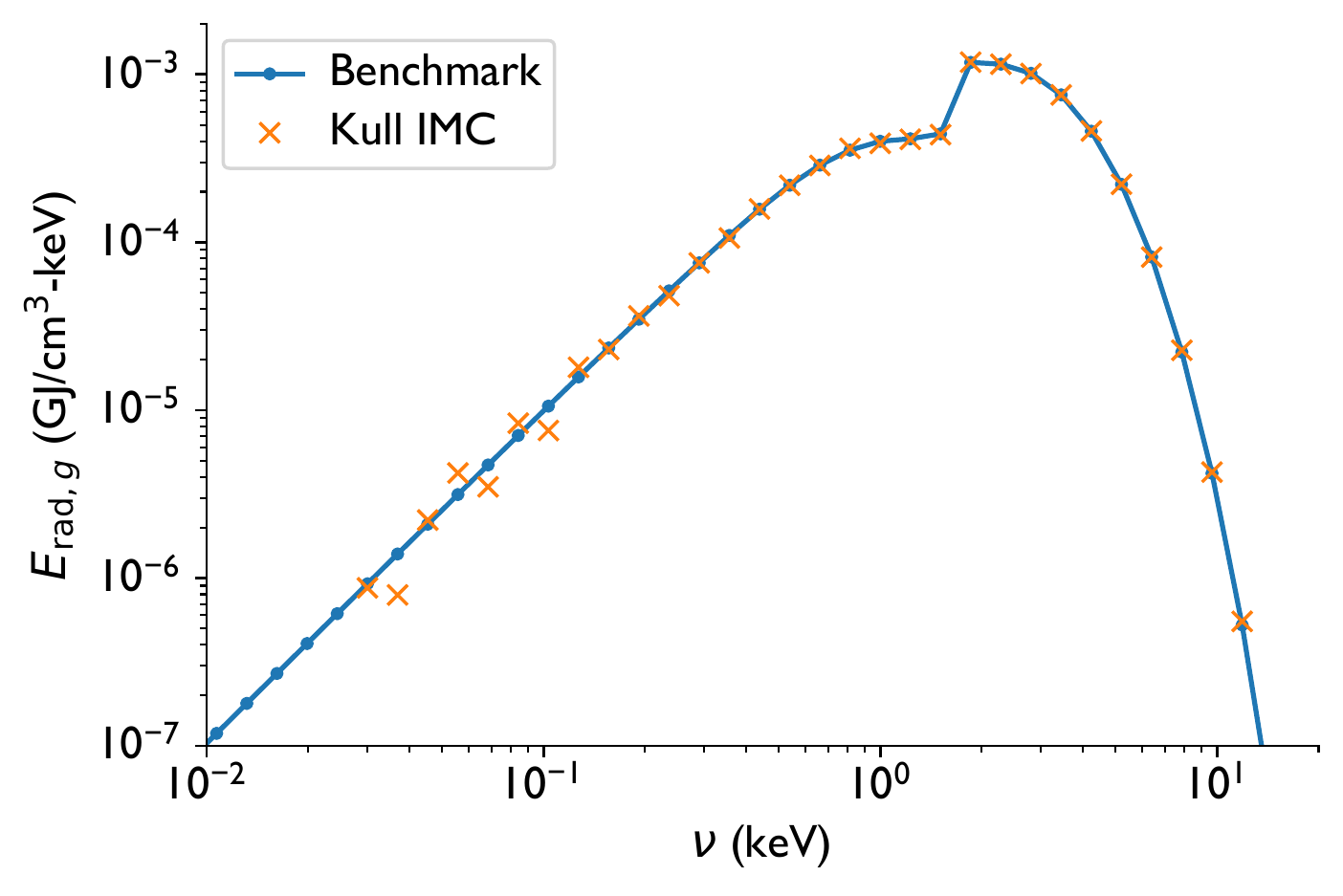}}
\subfigure[Absolute \% Error in Different Models]{\includegraphics[width=0.33\textwidth]{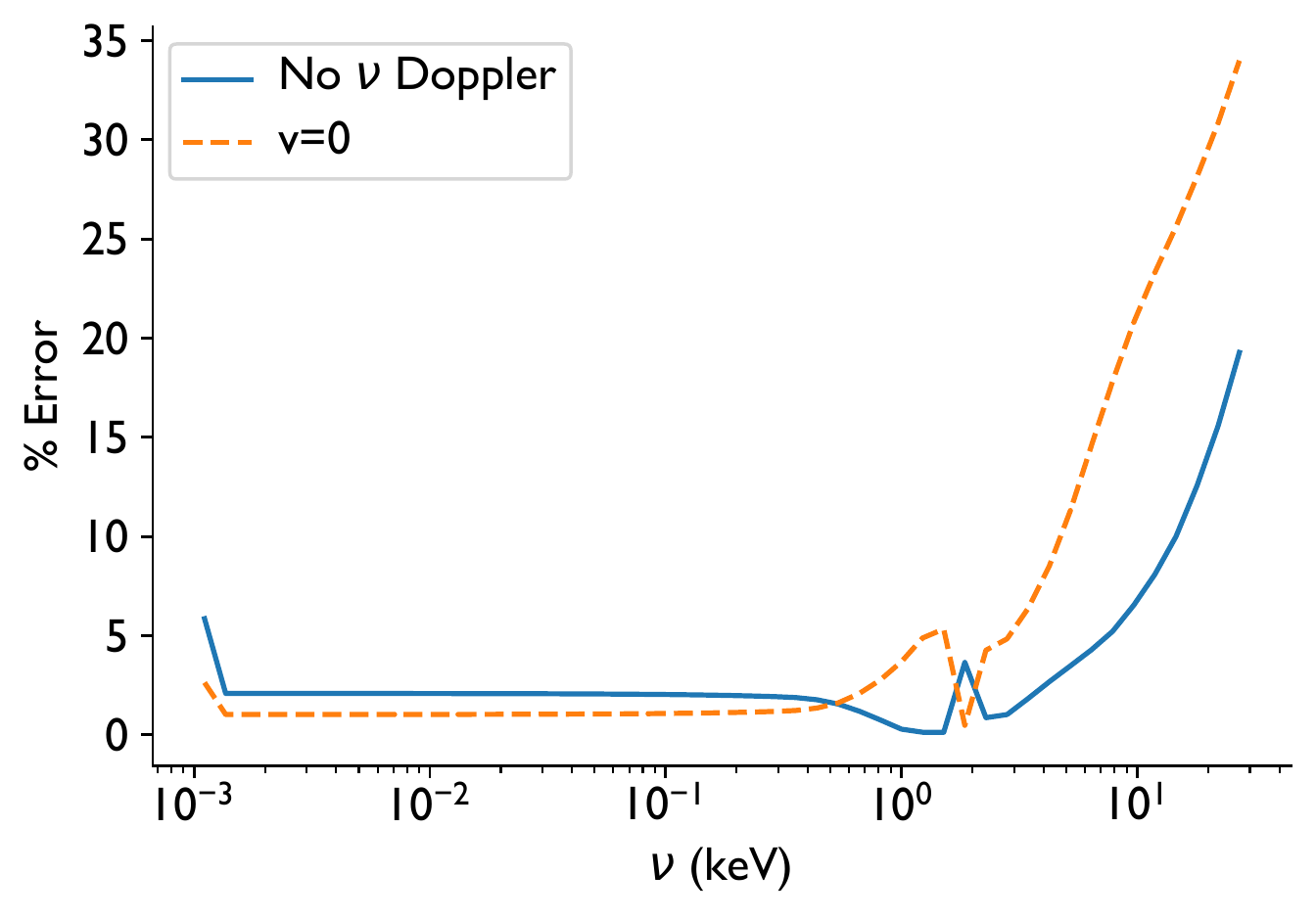}}
\subfigure[Absolute \% Error between KULL solutions with MMC and different analytic solutions]{\includegraphics[width=0.33\textwidth]{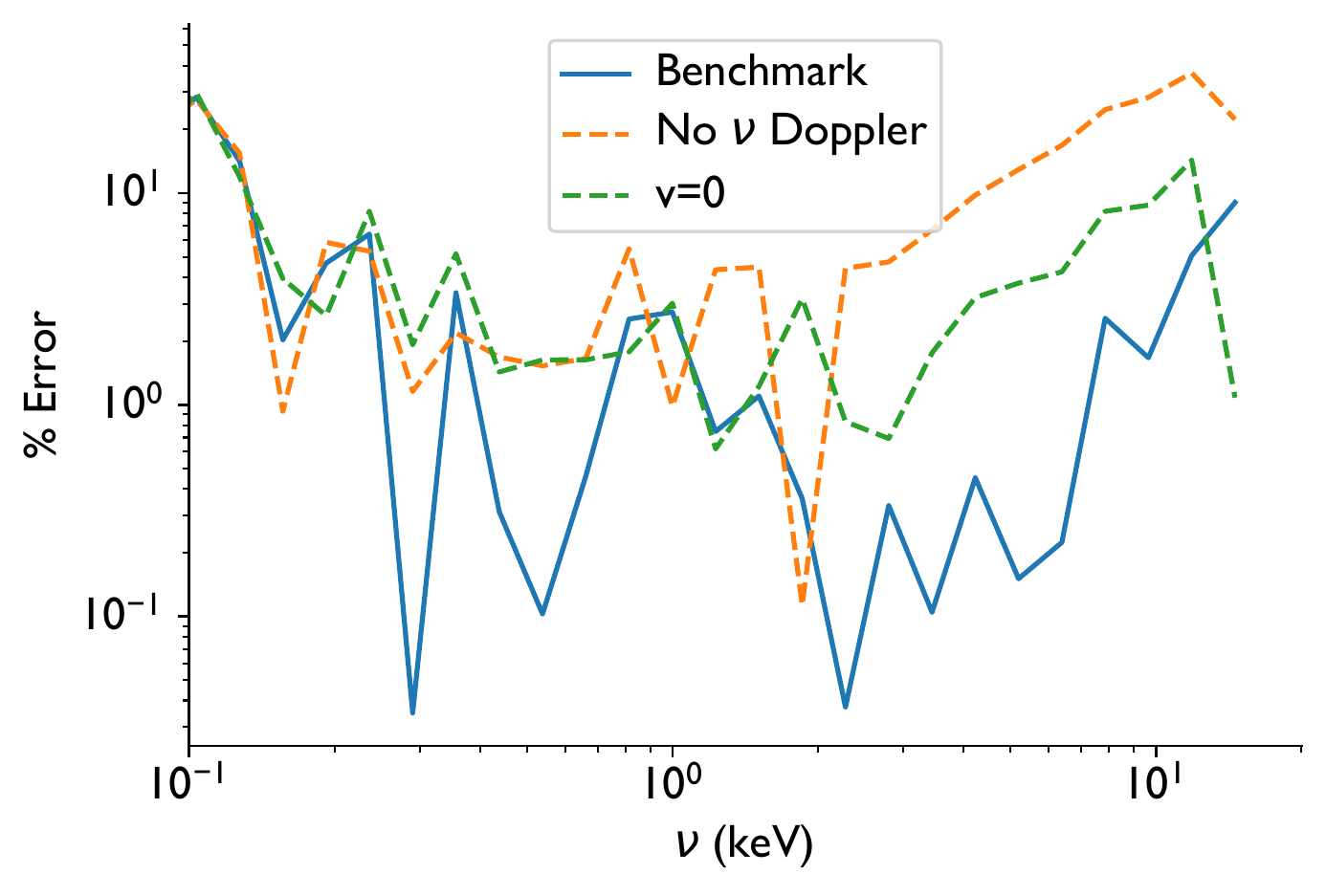}}
	\caption{ Comparisons for the 50 group aluminum slab problem: (a)  the benchmark solution and the KULL IMC solution with all MMC terms, (b) comparison of  the benchmark solution with a solution where the slab is stationary ($v=0$) and one where the Doppler correction for the frequency is ignored, and (c) comparison of the KULL IMC solution with full MMC with the benchmark solution, a solution with frequency Doppler shifts ignored, and with the slab stationary. }
	\label{fig:50gcomp}
\end{figure}

Results for the 50 group problem are shown in Figure \ref{fig:50gcomp}. In Figure \ref{fig:50gcomp}(a) the Kull solution with full MMC and $2\times 10^9$ particles per time step and the benchmark are compared, and reasonable agreement outside of the low-energy groups with few computational particles is observed.  The importance of MMC is quantified in Figure \ref{fig:50gcomp}(b) where the benchmark is compared with solutions where the slab is stationary and without the Doppler shift on $\nu$ via the percent absolute error. Here we can see that  errors up to 20\% occur in the higher energy groups when the Doppler shift of $\nu$ is ignored. Finally, Figure \ref{fig:50gcomp}(c) compares the Kull IMC solution with full MMC and the benchmark, as well as the solution lacking the Doppler correction and a solution where the slab is stationary. Figure \ref{fig:50gcomp}(c) indicates that above photon energies of 1 keV the Kull solution is, on the whole, closer to the benchmark than to the solutions missing MMC corrections.  We can conclude from this figure that our benchmark is able to verify if certain MMC terms are correctly implemented in a code.

\section*{ACKNOWLEDGEMENTS}

This work
    was performed under the auspices of the U.S. Department of Energy
    by Lawrence Livermore National Laboratory under Contract
    DE-AC52-07NA27344. Lawrence Livermore National Security, LLC. LLNL-ABS-813186

\newif\ifusebibtex
\usebibtexfalse

\setlength{\baselineskip}{12pt}
\bibliographystyle{mc2021}
\bibliography{mc2021}


\end{document}